\documentclass{aa}
\begin{document}

      \thesaurus{06     
              (08.12.1;  
               08.03.2;  
               08.16.4;  
               08.13.2;  
               08.02.4;  
               02.01.2 )}

\title{Orbital elements of barium stars
 formed through a wind accretion scenario}

\author{J. H. Liu$^{1,2}$, B. Zhang$^{1,5}$, Y. C. Liang$^3$, Q. H. Peng$^{4,5}$ 
       }
       
 \offprints{Liang Yanchun, email: lyc@yac.bao.ac.cn}   

\institute{  
  $^1$Department of Physics, Hebei Normal University, Shijiazhuang 050016, P.R. China\\
  $^2$Department of Physics, Shijianzhuang Teachers' College, Shijiazhuang 050041, P. R. 
      China\\
  $^3$Beijing Astronomical Observatory, Chinese Academy of Sciences, Beijing 100012, 
      P.R. China  \\
 $^4$Department of Astronomy, Nanjing University, Nanjing 210093, P. R. China \\
 $^5$Chinese Academy of Sciences-Peking University Joint Beijing Astrophysical Center, 
      Beijing 100871, P.R. China 
      }
      
\date{Received 3 July 2000 / Accepted 11 September 2000}
\maketitle

\markboth{J. H. Liu et al.: The orbital elements of barium stars}{}

\begin{abstract} 

Taking the total angular momentum conservation in place of the tangential 
momentum conservation, and considering the square and higher power terms
of orbital eccentricity $e$, the changes of orbital elements of binaries 
are recalculated for wind accretion scenario. These new equations of 
orbital elements are used to calculate the properties of barium stars. 
Our results show that, during the evolution of a binary system, the system 
widens as it loses mass, and the orbital period increases, while orbital 
eccentricity remains nearly constant, which can quantitatively explain the 
observed ($e$,log$P$) properties of normal G, K giants and those of barium 
stars. The results reflect the evolution from G, K giant binaries to barium 
binaries, namely, the orbits of barium stars have been modified by the
mass-transfer process responsible for their chemical peculiarities, whereas 
most of the G, K giant binaries are probably pre-mass transfer binaries.
Moreover, the results showed that the barium stars with longer orbital 
period $P$$>$1600 days may be formed by accreting part of the ejecta from 
the intrinsic AGB stars through wind accretion, while those with shorter 
orbital period may be formed through dynamically stable late case C mass 
transfer or common envelope ejection. 

\keywords{ Stars: late-type--  
                Stars: chemically peculiar--
                Stars: AGB--
                Stars: Mass loss--
                Binaries: spectroscopic
                }
\end{abstract}

%

\section{Introduction}

Extrinsic AGB stars include various classes of G- and K-type barium stars 
and those cooler S and C stars in which $~^{99}$Tc 
($\tau_{ 1\over 2}$=2$\times 10^5$ yr) is not observed. It is generally 
believed that they belong to binary systems and their heavy-element 
overabundance come from accretion of the mass ejected by their companions
(the former AGB stars, now evolved into white dwarfs) 
(McClure et al. 1980; Boffin \& Jorissen 1988; Jorissen \& Mayor 1992;
Jorissen \& Van Eck 2000; Jorissen et al. 1998; B\"{o}hm-Vitense et al. 2000; 
Karakas et al. 2000). The mass exchange took place about $1\times 10^6$ years 
ago, so the $^{99}$Tc produced in the original thermal pulse AGB 
(hereafter TP-AGB) stars have practically all decayed. 
The accretion may either be disk accretion (Iben $\&$ Tutukov 1985) or common 
envelope ejection (Paczynski 1976). Han et al. (1995) detailedly investigated 
the evolutionary channels for the formation of barium stars.

Boffin \& Jorissen (1988) calculated quantitatively the variations of orbital 
elements caused by wind accretion in binary systems. They also estimated the 
heavy-element overabundance of barium stars. In a subsequent paper 
(Boffin $\&$ Za$\check{c}$s 1994), the heavy-element overabundance of barium 
stars was calculated using the same technique, and the correlation with 
orbital period was interpreted. The calculated results were compared with the 
observations (Za$\check{c}$s 1994). In our calculation of the overabundance 
(Chang et al. 1997), we considered the variation in the binary separation, 
$\delta r \neq 0$, and combined the nucleosynthesis scenario of intrinsic 
TP-AGB stars (Zhang et al. 1998) with the scenario of successive pulsed 
accretion and mixing of wind mass.

Some important conclusions have been obtained in the theory of wind accretion,
but previous calculations on orbital elements were not very reliable
because of the neglect of the $\delta r/r$ term and the adoption of tangential 
momentum conservation (Chang et al. 1997 and references therein). For the 
rotating binary system with wind mass loss, total angular momentum conservation 
is a more realistic assumption than tangential momentum conservation.
Also, for the sake of simplicity, previous calculations regarded orbital 
eccentricity as $e\ll 1$ (Huang 1956; Boffin \& Jorissen 1988; Theuns et al. 1996). 
Observations show that the eccentricities of almost half of all barium stars 
are too large to be neglected, with the largest observed value = 0.97 
(Jorissen et al. 1998). Thus the assumption of $e\ll 1$ for the barium star 
systems is not valid.

The ($e$,log$P$) diagram is a very useful tool to study binary evolution given 
the abundant information contained in it. For example, the distributions of 
orbital period $P$ with the eccentricity of progenitor systems correlate with 
the final characteristics of binary systems through mass accretion. 
Generally, barium stars evolved from normal red giant binary systems. 
Observations show that the orbital eccentricities of Ba stars are smaller 
than those of the G, K giants from open clusters, and the ($e$,log$P$) diagrams 
of the two systems show strong correlation (Jorissen \& Boffin 1992; 
Boffin et al. 1993; Jorissen et al. 1998; Jorissen 1999).
Few quantitative calculations have been put into this correlation though they 
are very important to understanding the binary evolution.

In this paper, we adopt the hypothesis that the total angular momentum is 
conserved, and we do not neglect the square and the higher power terms of 
eccentricity, in order to recalculate the change equations of orbital elements of 
binary systems caused by stellar wind accretion. Then these new equations are used 
to calculate the properties of barium stars produced by this mechanism.
Hence we quantitatively explain the observed relations between the  
($e$,log$P$) diagram of normal G, K giants and that of barium stars.
In Sect. 2, we analyze the properties of ($e$,log$P$) diagrams of barium stars 
and normal G, K giants. In Sect. 3, we present the angular momentum conservation 
model of the wind accretion. Our results and analysis are illustrated in Sect. 4. 
And in Sect. 5, we draw some conclusions and discuss the results.

\section{The ($e$,log$P$) diagram}

The observations of the G, K giants from open cluster and barium stars are 
the same as in Jorissen et al. (1998), and are given in our Fig. 1 and 2
respectively. Analyzing the observations, we can understand the properties 
of them as following.

(1) At any given orbital period, the maximum eccentricity found among barium 
systems is much smaller than for cluster giants. And the average eccentricity 
of barium stars is generally smaller than that of cluster giants. The average 
orbital eccentricity of barium stars is 0.14, while the corresponding mean 
value of cluster giant is 0.23 (Boffin et al. 1993).

(2) In the two ($e$,log$P$) diagrams, the maximum eccentricity line of
barium system is almost parallel to the corresponding maximum eccentricity 
line of cluster giants. It seems that the maximum eccentricity line of 
cluster giants is moved in a parallel way with the increase of orbital period 
to reach the barium systems. Boffin et al. (1993) exhibited this character.
 
(3) In the ($e$,log$P$) diagrams, the average eccentricity line of barium 
stars is almost parallel to the corresponding line of cluster giants.
It seems that the latter is moved in a parallel way to the longer period 
region to reach the barium systems.

(4) In the ($e$,log$P$) diagram of normal G, K giants, the maximum orbital 
period for a circular orbit is $\sim$350 days; while the maximum orbital 
period for a circular orbit is $\sim$2000 days in the ($e$,log$P$) diagram of
barium stars, which is longer than the corresponding threshold of cluster 
giants. The ($e$,log$P$) diagrams of S and CH stars are very similar to those 
of barium stars (Jorissen et al. 1998). 
 
(5) Although, at a given orbital period, the maximum eccentricity found among 
barium systems is much smaller than for cluster giants. Still, the ($e$,log$P$) 
diagram of barium systems shows that quite large eccentricities are found
among barium stars, and the highest value is up to 0.97 (HD 123949), 
yet at large periods ($P$$\sim$9200 days for HD 123949).

The reasonable binary evolutionary model of orbital elements should
quantitatively explain the above mentioned observational properties.

\section{The angular momentum conservation model of wind accretion scenario}

For the binary system, the two components (an intrinsic AGB star with mass 
$M_{1}$, the present white dwarf star, and a main sequence star with mass 
$M_{2}$, the present barium star) rotating around the mass core C, so the 
total angular momentum is conserved in the mass core reference frame. If the 
two components exchange material through wind accretion, the angular momentum 
conservation of the total system is showed by:
\begin{equation}
\Delta(\mu r^{2}\dot{\theta})=\omega r_{1}^{2}(\Delta M_{1}+\Delta M_{2})
+r_{2}v(\Delta M_{1}+\Delta M_{2}),
\end{equation}
where $\mu$ is reduced mass, and $r$ is the distance from $M_{2}$ to $M_{1}$.
$r_{1}$, $r_{2}$ are the distances from $M_{1}$, $M_{2}$ to the mass core C 
respectively. $\omega$(=${2\pi}/P$, $P$=2$\pi A^2(1-e^2)^{1 \over 2}/h$, 
is orbital period) is angular velocity. $v$ is an additional effective velocity 
defined through the angular momentum variation in the direction of orbital 
motion of component 2. The first term on the right side of the equal-sign is 
the angular momentum lost by the escaping material, and the second term is the 
additional angular momentum lost by the escaping material.
 
Using the methods similar to those adopted by Huang (1956) and Boffin $\&$ 
Jorissen (1988), but considering the angular momentum conservation of the 
total system and not neglecting the square and higher power terms of 
eccentricity, we can obtain the formulas for the changes in the orbital 
elements:
\begin{eqnarray}
\frac{\Delta A}{A} & = &  - 2(1-e^2)^{1\over2} \left [ \frac{\Delta M_1}{M_1}+\frac{\Delta M_2}{M_2}
                 -  \frac {\Delta M_1+ \Delta M_2}{M_2} \frac {v} {v_{\rm orb}}\right ] \nonumber \\
                 & + & 2(1-e^2)^{1\over2} \frac {M_2 (\Delta M_1+ \Delta M_2)}{M_1(M_1+M_2)}  \nonumber \\
                & + & [2(1-e^2)^{1\over2}-1] \frac {\Delta M_1+ \Delta M_2}{M_1+M_2},            
\end{eqnarray}

\begin{eqnarray}
\frac{e\Delta e}{1-e^2} & = &  [1-(1-e^2)^ {1\over2}]
                \left [ \frac {\Delta M_1}{M_1}+\frac {\Delta M_2}{M_2}         
                - \frac {\Delta M_1+ \Delta M_2}{M_1+M_2} \right ] \nonumber \\
              & - & [1-(1-e^2)^ {1\over2}]\frac {M_2 (\Delta M_1+ \Delta M_2)}{M_1(M_1+M_2)}  \nonumber \\
              & - & \frac {3e^2}{2(1-e^2)^{1\over2}} \frac {\Delta M_1+ \Delta M_2}{M_2}
              \frac {v}{v_{\rm orb}} ,
 \end{eqnarray}
where $A$ is the semi-major axis of the relative orbit of component 2 around 1, 
and $e$ is the orbital eccentricity (more details can be found in Appendix).  
Here, we take $v=0$ (Boffin $\&$ Jorissen 1988). 

For the mass being accreted by the barium star, we use the Bondi-Hoyle accretion rate 
(Theuns et al. 1996; Jorissen et al. 1998):
\begin{equation}
\Delta M_{2}^{\rm acc}=- \frac {\alpha}{A^{2}}
                     \left [\frac{GM_{2}}{v_{\rm ej}^{2}} \right ]^{2}
     \left [\frac{1} {1+(v_{\rm orb}/v_{\rm ej})^{2}} \right ]^{3\over2} 
     \Delta M_{1} ,
\end{equation}
where $\alpha$ is a constant expressing the accretion efficiency, 
and $\alpha$$\sim$0.05 in the situation of interest here according
to the detailed hydrodynamic simulations (Theuns et al. 1996;
Jorissen et al. 1998). $e$ is the orbital eccentricity. $v_{\rm ej}$ is the wind 
velocity and $v_{\rm orb}$ is the orbital mean velocity. After fixing the initial 
conditions, for the mass $\Delta M_1$, ejected at each pulse by the primary star, 
we can solve the Eqs. (2)$-$(4) for $\Delta M_2$, the mass accreted by the secondary 
star. The final mass and critical Roche lobe radii of the intrinsic AGB stars,
and the initial envelope mass of the extrinsic AGB stars are all calculated using 
the same formulas as in Boffin \& Jorissen (1988) in which the mass ratio is 
$q=M_1/M_2$.

\section{Results and analysis}

Adopting the wind accretion scenario above mentioned, we calculate the orbital
elements of barium systems. We take as standard case: $M_{1,0}$=3.0 $M_{\odot}$,
$M_{2,0}$=1.3$M_{\odot}$, $v_{\rm ej}$=15$~$km$~$s$^{-1}$ 
(Boffin $\&$ Za$\check{c}$s 1994). The results are showed in Figs. 1$-$4 and Table 1.

\begin{figure}
\input epsf
\epsfverbosetrue
\epsfxsize 8.8cm 
\epsfbox{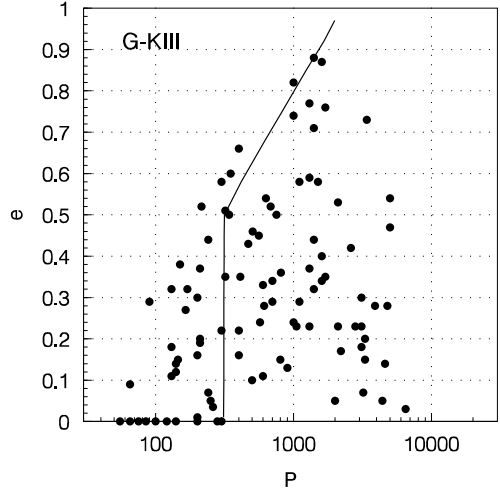}  
\caption{~The ($e$,log$P$) diagram for the G and K giants. The solid 
represents the maximum orbital eccentricity line of red giant binaries.       } 
\end{figure}

\begin{figure}
\input epsf
\epsfverbosetrue
\epsfxsize 8.8cm 
\epsfbox{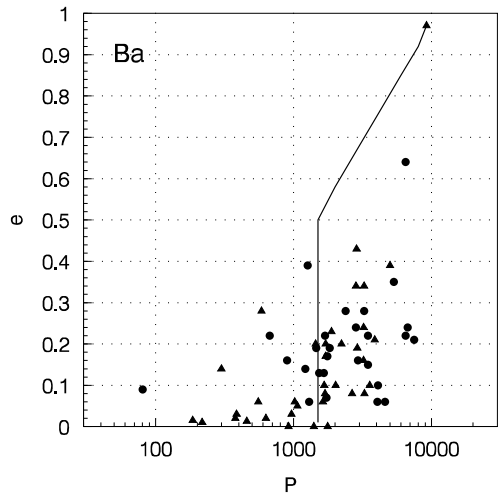} 
\caption{~The ($e$,log$P$) diagram for barium stars. The solid 
represents the calculated maximum orbital eccentricity of barium star 
systems created by the wind accretion scenario.}

\end{figure}

\begin{figure}
\input epsf
\epsfverbosetrue
\epsfxsize 8.8cm
 \epsfbox{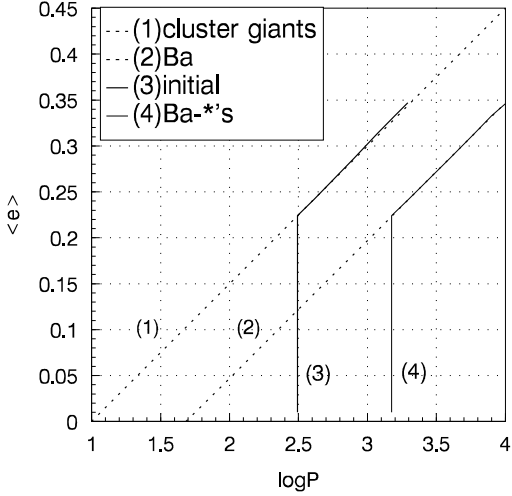} 
\caption{~Mean eccentricity$-$log$P$ diagram for binary systems. 
The dashed line (1) represents the fit of mean eccentricity of observed 
cluster giants taken from Boffin et al. (1993), and the dashed line (2) 
represents the corresponding fit of mean eccentricity of observed Ba stars.
The thin solid line (4) represents the orbital eccentricity of barium stars 
with $\eta=0.685$ in Eq. (11) of Boffin et al. (1993), and the thick solid 
line (3) represents the corresponding initial eccentricity of binary systems. }
\end{figure}

\begin{table} 
\begin{flushleft} 
{\bf Table 1.} Variations of the eccentricity $e$ during the process of mass 
loss by wind as a function of the mass ratio $q$ for the different initial 
eccentricities. The initial eccentricities are 0.1, 0.3, 0.5 and 0.75 
respectively. 
\end{flushleft}
\begin{center}  
\begin{tabular}{ccccc} \hline

  $q$   &       &    $e$   &    &            \\  \hline
       & $e_0=0.1$ &   $e_0=0.3$  &  $e_0=0.5$ &  $e_0=0.75$    \\
 2.31  &  0.1    &  0.3    &   0.5     &  0.75          \\ 
 2.21  & 0.1001  &  0.3002 &  0.5003   & 0.7503            \\
 2.11  & 0.1001  &  0.3004 &  0.5007   & 0.7506              \\ 
 2.01  & 0.1002  &  0.3006 &  0.5010   & 0.7508              \\
 1.91  & 0.1003  &  0.3008 &  0.5013   & 0.7511      \\
 1.81  & 0.1004  &  0.3010 &  0.5016   & 0.7513       \\
 1.72  & 0.1004  &  0.3012 &  0.5019   & 0.7516        \\
 1.62  & 0.1005  &  0.3014 &  0.5022   & 0.7518     \\
 1.52  & 0.1005  &  0.3016 &  0.5025   & 0.7520       \\
 1.43  & 0.1006  &  0.3017 &  0.5027   & 0.7522        \\
 1.33  & 0.1006  &  0.3019 &  0.5029   & 0.7524      \\
 1.24  & 0.1007  &  0.3020 &  0.5031   & 0.7525      \\
 1.14  & 0.1007  &  0.3020 &  0.5032   & 0.7526     \\
 1.05  & 0.1007  &  0.3021 &  0.5033   & 0.7527       \\
 1.00  & 0.1007  &  0.3021 &  0.5033   & 0.7527     \\
 0.86  & 0.1007  &  0.3020 &  0.5032   & 0.7526     \\
 0.77  & 0.1006  &  0.3019 &  0.5030   & 0.7525      \\
 0.68  & 0.1006  &  0.3017 &  0.5026   & 0.7522     \\
 0.59  & 0.1005  &  0.3013 &  0.5020   & 0.7519      \\
 0.50  & 0.1003  &  0.3008 &  0.5012   & 0.7513     \\
 0.41  & 0.1000  &  0.3001 &  0.5000   & 0.7506      \\ \hline 

\end{tabular}
\end{center}
\end{table} 

\begin{figure}
\input epsf
\epsfverbosetrue
\epsfxsize 8.8cm   
 \epsfbox{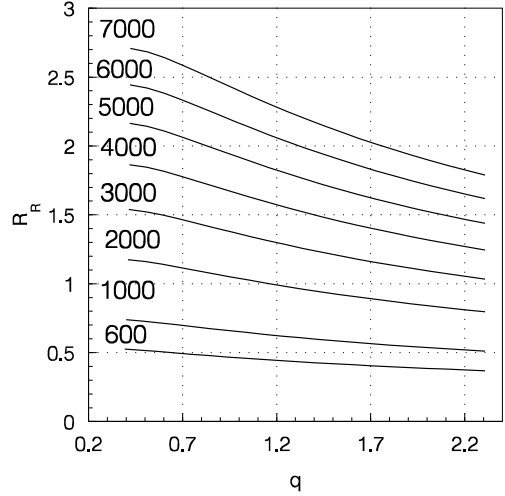}     
\caption{~Evolution of the Roche lobe radius $R_R$ in AU 
(solid lines) for the different final orbital periods during the process
of mass loss by wind as a function of the mass ratio $q$. The final 
values of orbital periods (in days) are indicated.}
\end{figure}

According to the suggestions of Jorissen et al. (1998) and Zhang et al. (1999),
the barium stars with longer orbital periods ($P$$>$1500 d and $P$$>$1600 d 
respectively) are formed through stellar wind accretion scenario. So we adopt 
1600 days to be the lower bound of wind accretion for the formation of barium 
stars in our calculation. Fig. 1 show the maximum eccentricities line of the 
normal giants (solid line), which is regarded as the initial orbit of the binary 
system. The observations are for the G, K giants in open cluster 
(Mermilliod 1996; Jorissen et al. 1998). While the solid line in Fig. 2 represents 
the calculated eccentricity of barium stars through our wind accretion model,
which is near to the observed maximum eccentricities line of barium stars.

Fig. 3 show the ($e$,log$P$) results about mean orbital eccentricities. The dashed 
line (1) represents the fit of mean eccentricity of observed cluster giants,
which is taken from Boffin et al. (1993); the dashed line (2) represents the fit 
of mean eccentricity of observed barium stars taken from Boffin et al. (1993).
The thin solid line (4) represents the orbital eccentricity of barium stars with 
$\eta=0.685$ in Eq. (11) of Boffin et al. (1993) (Boffin \& Jorissen (1988) 
predicted values of $\eta$ in the range 0.3 to about 1). In calculation, 
we firstly fix the eccentricity of barium stars at given periods 
(thin solid line (4)), then use our wind accretion model to determine the 
corresponding initial eccentricity of binary systems (thick solid line (3)), 
which nearly overlap the observed mean values of cluster giants (dashed line (1)). 

These results for the ($e$,log$P$) diagrams show that the barium stars indeed 
evolve from the G, K giants, and the properties of barium stars with longer 
orbital periods ($P$$>$1600 d) can be explained by the wind accretion scenario.

Table 1 reports the variations of the eccentricities $e$ during the process of 
mass loss by wind as a function of mass ratio $q$ ($q$=$M_1/M_2$) for different
initial eccentricities of binary systems. The initial eccentricities are 
0.1, 0.3, 0.5 and 0.75 respectively. The results show that $e$ is very nearly 
constant in the course of the wind accretion, which is different from the 
result of Theuns et al. (1996).
 
Fig. 4 shows the evolution of the Roche radius due to mass loss via the wind  
as a function of the mass ratio $q$=$M_1/M_2$ for different final orbital periods 
(in unit of days) of barium systems. The results show that, for a given final 
period of a barium star, the mass loss through stellar wind leads to the 
increase of the Roche radius, thus making Roche lobe overflow (hereafter RLOF) 
more difficult to achieve. Moreover, the longer final orbital period, the 
larger Roche radius will be obtained, hence it is more difficult to occur the 
Roche lobe overflow scenario, so that the circular orbit forms with more 
difficulties.
 
\section{Conclusions and discussion}

Taking the conservation of angular momentum in place of the conservation of 
tangential momentum for wind accretion scenario, and considering the square 
and higher power terms of orbital eccentricity, the change equations of orbital
elements have been recalculated. These equations have been applied to 
quantitatively calculate the orbital elements of barium stars and to discuss
the stellar wind accretion scenario.

Our results show that during the wind accretion process, the system widens 
as it loses mass, the orbital semi-major axis $A$ will increase, resulting 
in an increase in the orbital period, which is similar to the result of 
Boffin \& Jorissen (1988), but the eccentricity is very nearly constant.
In general, barium stars evolved from normal G, K giants, namely, barium stars 
formed by accreting the ejecta of the G, K giant companions of binary systems. 
If the accretion is through stellar wind, the orbital period will increase and 
eccentricity remains nearly constant. Thus, for the same eccentricity, the 
orbital periods of barium stars are longer than those of normal G, K giants; 
while for the same orbital period, the maximum and average eccentricity
of barium stars are lower than the corresponding values of the normal G, K 
giants. Our results can explain quantitatively these relations. Namely,
we can explain the properties of ($e$,log$P$) diagrams of normal cluster 
giants and barium stars shown in Sect. 2.

(1) Eq. (2) represents the change of orbital semi-major axis $A$, which shows 
that $A$ will increase during the wind accretion process, resulting the 
increase of orbital periods. But the orbital eccentricity will be very nearly 
constant with the wind accretion (see Eq. (3) and Table 1). These results 
quantitatively explain the observational fact that the maximum eccentricities 
of barium stars is lower than the corresponding maximum eccentricities of 
normal giants at any given orbital periods.

(2) Regarding the maximum eccentricity line of normal G, K giants as the 
initial condition of a binary system, the  maximum eccentricity line of barium 
stars can be obtained through the wind accretion calculation, which can fit the 
observations. Moreover, the two maximum eccentricity lines are near parallel 
(see Fig. 1 and 2). 

(3) The average orbital eccentricity line of barium stars can be obtained 
through the wind accretion calculation, and can fit the observations.
Namely, adopting $\eta=0.685$ for Eq. (11) of Boffin et al. (1993) to
obtain the average eccentricities of barium stars, the initial eccentricities 
of binary systems can be obtained using our wind accretion model, which nearly 
overlap the observed average values of normal G, K giants given by Boffin et al. 
(1993) (see Fig. 3). 

(4) For the low eccentricity systems, the observed maximum orbital period of 
circular orbits is $\sim$350 days for the cluster giants. Our result shows that 
this upper bound will lead to the corresponding maximum circular orbital period
near 2000 days for barium stars using the wind accretion model, which can fit 
the observations. The threshold periods of CH and S stars are very similar to 
that of barium stars.

(5) Taking the highest orbital eccentricity 0.9 of normal G, K giants as the 
initial condition, we can calculate a large eccentricity, up to 0.9, for barium 
stars. The reason may be that the orbital period increases, but the eccentricity 
is nearly constant during the wind accretion process, so that the long orbital 
period and high eccentricity barium systems are formed. Our results can 
quantitatively explain the observed large eccentricity, 0.9, of some
barium systems.
                                
These fits between our results and the corresponding observations show that
barium systems evolved from the normal G, K giants, and the distribution of
orbital eccentricity with orbital period of barium stars can reflect
the final orbital properties of binary systems through wind accretion 
for the binaries with longer period ($P$$>$1600 d).

Moreover, this period value can be understood roughly by the results showed 
in Fig. 4. Fig. 4 shows the evolution of Roche radius as a function of mass 
ratio $q=M_1/M_2$. The wind accretion scenario needs the larger Roche radius 
than the stellar radii of the previous primary stars of the binary systems, 
AGB stars. If we also consider the radii of AGB stars:
a typical AGB star with log$T_{\rm eff}=3.5$ and $M_{\rm bol}=-5$ has a radius 
of 280$R_{\odot}$ (Jorissen \& Boffin 1992), the comparison between the evolution 
of the radii of the AGB stars and the evolution of the Roche radii will show that 
the Roche radii are larger than the corresponding stellar radii for the binary 
systems with longer final orbital period ($P$$>$1600 d) during the binary 
evolutions. This result reflect that the barium stars with longer orbital period 
($P$$>$1600 d) form through the wind accretion scenario. 

In addition, Liang et al. (2000) confirmed this orbital period value taking 
advantage of the observational heavy element abundances of barium stars.

For the barium systems with shorter orbital periods, our wind 
accretion model is not sufficient to explain the observed properties.
They may be formed through dynamical stability case C Roche lobe overflow,
or common envelope ejection, or some other scenarios 
(Jorissen et al. 1995; Jorissen et al. 1998;
Han et al. 1995 and references therein). Among the binary systems with shorter 
periods, the primary star may fill the Roche radius when it evolve to AGB phase,
thus the mass-transfer process occur, but the previously strong wind mass loss 
has made the mass of primary star become very small so that mass ratio
$q$ ($=M_1/M_2$) is lower than a threshold value $q_{\rm crit}$ 
($\sim$0.65, Pastetter \& Ritter 1989), which causes the dynamically stable RLOF. 
While if the mass ratio $q$ is greater than the $q_{\rm crit}$ when the primary 
AGB star fills its Roche lobe, the  Roche lobe overflow will be dynamically 
unstable, leading to the formation of common envelope. Friction between the 
stars and the common envelope causes contraction of the binary orbit and release 
of energy. Most of the energy is used in the ejection of the common envelope. 
The binary after the ejection of common envelope comprises a white dwarf and a 
barium star with accreted neutron capture process elements$-$rich material. 
For such a binary, the orbital period and eccentricity are lower. The barium star 
HD 77247 ($P=80.53$ d, $e=0.09$) can be explained by the common envelope scenario.

\begin{acknowledgements}
We thank the anonymous referee for very useful suggestions on the original 
manuscript. We thank Dr. Alain Jorissen for fruitful discussion and 
warm-hearted help. Thank Dr. John Lattanzio for sending important material to 
us. This research work is supported by the National Natural Science Foundation 
of China under grant No. 19973002.
\end{acknowledgements}

\section * {\bf Appendix: derivation of Eqs. (2) and (3)}

For the binary system, the two components (an intrinsic AGB star with mass $M_{1}$, 
the present white dwarf star, and a main sequence star with mass $M_{2}$,
the present barium star) rotating around the mass core C, so the total angular 
momentum is conservative in the mass core reference frame. If the mass exchange
between the two components is wind accretion, the total angular momentum
conservation is shown by:
\begin{equation}
\Delta(\mu r^{2}\dot{\theta})=\omega r_{1}^{2}(\Delta M_{1}+\Delta M_{2})
+r_{2}v(\Delta M_{1}+\Delta M_{2}),
\end{equation}
where $\mu$ is reduced mass, and $r$ is the distance from $M_{2}$ to $M_{1}$.
$r_{1}$ and $r_{2}$ are the distances from $M_{1}$, $M_{2}$ to the mass core C 
respectively. $\omega$ (={2$\pi /P$) is angular velocity, where 
$P$=2$\pi A^2(1-e^2)^{1 \over 2}/h$ is the orbital period (Huang 1956).  
$v$ is an additional effective velocity defined through the angular momentum 
variation in the direction of orbital motion of component 2. The first term 
on the right side of the equal-sign is the angular momentum lost by the 
escaping material and the second term is the additional angular momentum 
lost by the escaping material.

For the binary system, according to Huang (1956), the changes of
orbital elements, the orbital semi-major axis $A$ and eccentricity $e$, are

\begin{equation}
\frac {\Delta A}{A}=\frac {\Delta(M_1+M_2)}{M_1+M_2}-\frac {\Delta E}{E},
\end{equation}
\begin{equation}
\frac {e\Delta e}{1-e^2}=\frac {\Delta(M_1+M_2)}{M_1+M_2}-
{1\over2} \frac {\Delta E}{E}-\frac {\Delta h}{h},
\end{equation}

where
\begin{eqnarray}
\frac {\Delta E}{E} & = & \frac {\Delta T}{E}+\frac {\Delta \Omega}{E}  \nonumber\\
    & = & \frac {r\dot {\theta}}{E} \Delta (r\dot {\theta})
          + \left [\frac {\Delta (M_{1}+M_{2})}{M_{1}+M_{2}}\frac{2A}{r}
          -\frac {2A\Delta r}{r^2} \right ]~.  
\end{eqnarray}

According to the angular momentum conservation model, 
the $\Delta (r\dot {\theta})$ term can be obtained from the following 
equation:
\begin{eqnarray}
\frac{M_1M_2}{M_1+M_2}r\Delta (r\dot {\theta}) 
            & = & \Delta \left ({\frac{M_1M_2}{M_1+M_2}r~r\dot {\theta}} \right ) \nonumber\\
            & - & \Delta \left ({\frac{M_1M_2}{M_1+M_2}r} \right )~r\dot {\theta}  \nonumber\\
            & = & \Delta(\mu r^{2}\dot{\theta})
                  -\Delta \left ({\frac{M_1M_2}{M_1+M_2}r} \right )~r\dot {\theta}    \nonumber\\
            & = &  \omega r_{1}^{2} (\Delta M_{1}+\Delta M_{2})            \nonumber\\
            & + & r_{2}v(\Delta M_{1}+\Delta M_{2})                      \nonumber\\
            & - & \Delta \left ({\frac{M_1M_2}{M_1+M_2}r} \right )~r\dot {\theta}, 
\end{eqnarray}

thus
\begin{eqnarray}
\Delta (r\dot {\theta}) & = & -r\dot {\theta} \left [\frac {\Delta {M_1}}{M_1}
                          +\frac {\Delta {M_2}}{M_2}
                          -\frac {\Delta (M_1+M_2)}{M_1+M_2} \right ]
                          -r\dot {\theta} \frac {\Delta r}{r} \nonumber\\
                    & + & \frac {\Delta (M_{1}+M_{2})}{M_{2}}v   
                         +\frac {rG^ {1\over2}M_{2}\Delta (M_{1}+M_{2})}
                          {M_{1}(M_{1}+M_{2})^ {1\over2}A^{3\over2}}~,
\end{eqnarray}

then, we can obtain
\begin{eqnarray}
\frac {r\dot {\theta}\Delta (r\dot {\theta})}{E} 
           & = & -\frac {(r\dot {\theta})^2}{E} \left [\frac {\Delta {M_1}}{M_1}
                          +\frac {\Delta {M_2}}{M_2}
                          -\frac {\Delta (M_1+M_2)}{M_1+M_2} \right ] \nonumber\\
                  & - & \frac {r\dot {\theta}^2}{E} \Delta r 
                          +\frac {r\dot {\theta}}{E}\frac {\Delta (M_{1}+M_{2})}{M_{2}}v   \nonumber\\
                  & + &  \frac {r^2\dot {\theta}}{E}\frac {G^ {1\over2}M_{2}\Delta (M_{1}+M_{2})}
                          {M_{1}(M_{1}+M_{2})^ {1\over2}A^ {3\over2}}    \nonumber\\
           & = & 2(1-e^2)^ {1\over2}
                      \left [\frac {\Delta {M_1}}{M_1}+\frac {\Delta {M_2}}{M_2}                                            
                       -\frac {\Delta (M_{1}+M_{2})}{M_2} \frac {v} {v_{\rm orb}} \right ] \nonumber\\
                  & - & 2(1-e^2)^ {1\over2}\frac {M_2 \Delta (M_1+ M_2)}{M_1(M_1+M_2)} \nonumber\\
                  & - & 2(1-e^2)^ {1\over2}\frac {\Delta(M_1+M_2)}{M_1+M_2}
                     +\frac {h\dot {\theta}\Delta r}{Er}    \nonumber\\            
           & = & 2(1-e^2)^ {1\over2}
                      \left [\frac {\Delta {M_1}}{M_1}+\frac {\Delta {M_2}}{M_2}                                            
                       -\frac {\Delta (M_{1}+M_{2})}{M_2} \frac {v} {v_{\rm orb}} \right ] \nonumber\\
                  & - & 2(1-e^2)^ {1\over2}\frac {M_2 \Delta (M_1+ M_2)}{M_1(M_1+M_2)}  \nonumber\\
                  & - & 2(1-e^2)^ {1\over2}\frac {\Delta(M_1+M_2)}{M_1+M_2}        
                      +\frac {2\Delta r}{A(1-e^2)^{1\over2}}~.
\end{eqnarray}

Thus
\begin{eqnarray}
\frac {\Delta E}{E}  & = & 2(1-e^2)^ {1\over2}         
                      \left [\frac {\Delta M_1}{M_1}+\frac {\Delta M_2}{M_2}
                       -\frac {\Delta (M_{1}+M_{2})}{M_2} \frac {v} {v_{\rm orb}} \right ] \nonumber\\ 
                  & - & 2(1-e^2)^ {1\over2} \frac {M_2 \Delta (M_1+ M_2)}{M_1(M_1+M_2)}    \nonumber\\                  
                  & + & (2-2(1-e^2)^ {1\over2})\frac {\Delta(M_1+M_2)}{M_1+M_2} ~.        
\end{eqnarray}

The $\Delta h/h$ term can be obtained from:
\begin{eqnarray}
\Delta(\mu r^2 \dot \theta) & = & \Delta({\mu}{h})           \nonumber\\  
              & = & \Delta(\frac{{M_1}{M_2}}{M_1+M_2}h)     \nonumber\\  
              & = & \Delta (\frac{{M_1}{M_2}}{M_1+M_2}) h+\frac{{M_1}{M_2}}{M_1+M_2}\Delta{h} \nonumber\\
              & = & {\omega} r_{1}^2 \Delta{(M_1+M_2)}+r_{2}v\Delta{(M_1+M_2)}~,
\end{eqnarray}
and
\begin{eqnarray}
\frac {\Delta h}{h} & = & -\frac {\Delta M_{1}}{M_{1}}-\frac {\Delta M_{2}}{M_{2}}                                                  
                     +  \frac {2+e^2}{2(1-e^2)^{1\over2}} \frac {\Delta (M_1+ M_2)}{M_{2}}
                     {v\over v_{\rm orb}}                       \nonumber\\  
          & + & \frac {M_2 \Delta (M_{1}+M_{2})}{M_{1}(M_{1}+M_{2})} 
           +\frac {\Delta (M_{1}+M_{2})}{M_{1}+M_{2}}  ~. 
\end{eqnarray}

Combining Eqs. (6), (7), (12) and (14), we can obtain the changes of orbital 
semi-major axis $A$ and eccentricity $e$:

\begin{eqnarray}
\frac{\Delta A}{A} & = &  - 2(1-e^2)^{1\over2} 
                 \left [\frac{\Delta M_1}{M_1}+\frac{\Delta M_2}{M_2}
                 -  \frac {\Delta M_1+ \Delta M_2}{M_2} \frac {v} {v_{\rm orb}} \right ] \nonumber \\
                & + & 2(1-e^2)^{1\over2}\frac {M_2 (\Delta M_1+ \Delta M_2)}{M_1(M_1+M_2)}  \nonumber \\
                & + & [2(1-e^2)^{1\over2}-1] \frac {\Delta M_1+ \Delta M_2}{M_1+M_2},            
\end{eqnarray}

\begin{eqnarray}
\frac{e\Delta e}{1-e^2} & = &  [1-(1-e^2)^ {1\over2}]
               \left [\frac {\Delta M_1}{M_1}+\frac {\Delta M_2}{M_2}         
                - \frac {\Delta M_1+ \Delta M_2}{M_1+M_2} \right ] \nonumber \\
              & - & [1-(1-e^2)^ {1\over2}] \frac {M_2 (\Delta M_1+ \Delta M_2)}{M_1(M_1+M_2)}  \nonumber \\
              & - & \frac {3e^2}{2(1-e^2)^{1\over2}} \frac {\Delta
              M_1+ \Delta M_2}{M_2} \frac {v}{v_{\rm orb}}.
\end{eqnarray}

For obtaining the above equations, averaging over the orbital
period then amounts to using the following mean values: 
\begin{eqnarray*}
<\frac{\dot \theta}{r}>=\frac{2\pi}{PA(1-e^2)} 
\end{eqnarray*}
\begin{eqnarray*}
<r>=A(1+{1\over 2} e^2)
\end{eqnarray*}
\begin{eqnarray*}
<r\dot \theta>=\frac{2 \pi A(1-e^2)^{1\over2}}{P}
\end{eqnarray*}
\begin{eqnarray*}
<\dot \theta>=\frac{2\pi}{P}
\end{eqnarray*}
\begin{eqnarray*}
<(r\dot \theta)^2>=(\frac{2 \pi A}{P})^2(1-e^2)^{1\over2}
\end{eqnarray*}
\begin{eqnarray*}
<\frac{1}{r}>=\frac{1}{A}
\end{eqnarray*}
\begin{eqnarray*}
<\frac{1}{r^2}>=\frac{1}{A^2(1-e^2)^{1\over2}}
\end{eqnarray*}

\end{document}